\documentclass[10pt,aps,prd,twocolumn,showpacs,superscriptaddress,nofootinbib,nobibnotes,longbibliography,floatfix]{revtex4-1}

\usepackage[utf8]{inputenc}
\usepackage[T1]{fontenc}
\usepackage{comment}
\usepackage{bm}
\usepackage{mathtools,amsmath,amssymb,amsfonts,mathrsfs,eucal,graphicx, tensor,csquotes, accents, commath, chngcntr, siunitx}
\usepackage[dvipsnames]{xcolor}
\usepackage[unicode]{hyperref}
\hypersetup{colorlinks=true, citecolor=MidnightBlue,
            linkcolor=MidnightBlue, urlcolor=WildStrawberry, linktocpage=true}
\usepackage[normalem]{ulem}

\DeclareMathAlphabet{\mathpzc}{OT1}{pzc}{m}{it}
\definecolor{darkgreen}{rgb}{0.0, 0.6, 0.0}
\labelformat{section}{Section #1} 
\labelformat{subsection}{Section #1} 
\labelformat{subsubsection}{Section #1}
\labelformat{subsubsubsection}{Section #1}
\labelformat{equation}{Eq.~(#1)} \labelformat{figure}{Fig.~#1} 
\labelformat{subfigure}{Fig.~\thefigure#1} 
\labelformat{table}{Table.~#1} 
\labelformat{appendix}{Appendix #1}
\numberwithin{equation}{section}

\begin{document}
\title{Black Hole Spectroscopy with Conditional Variational Autoencoder}

\author{Akash K Mishra}
\email{akash.mishra@saha.ac.in}
\affiliation{Theory Division, Saha Institute of Nuclear Physics, 1/AF Bidhan Nagar, Kolkata 700064, India}

\begin{abstract}
Gravitational waves provide a unique opportunity to test general relativity in the strong-field regime, enabling the extraction of key physical parameters from observational data. Traditional likelihood-based inference methods, while robust, become computationally expensive in high-dimensional parameter spaces, such as when incorporating multiple ringdown modes or beyond Kerr deviations. In this paper, we explore the implementation of a conditional variational autoencoder-based machine-learning framework for accelerated ringdown parameter estimation. As a first application, we use the neural network to infer the remnant properties of a final black hole under the Kerr hypothesis. 
We demonstrate the performance of this algorithm with simulated ringdown waveforms consistent with advanced LIGO sensitivity and compare with Bayesian analysis results. We further extend the framework beyond the Kerr paradigm by incorporating deviations predicted in braneworld gravity.

\end{abstract}
\maketitle
\section{Introduction}\label{sect:intro}
The past decade of gravitational wave observation by LIGO and Virgo~\cite{LIGOScientific:2016aoc, LIGOScientific:2016vlm, LIGOScientific:2020ibl, KAGRA:2021vkt} has revolutionized our understanding of strong field gravity and opened up unprecedented research avenues. This has provided crucial insights into some of the most complex astrophysical phenomena, validating the predictions of general relativity in highly non-linear regimes~\cite{LIGOScientific:2016lio, LIGOScientific:2021sio, LIGOScientific:2017vwq}. Additionally, these observations now provide direct measurements of component masses, individual spins, orbital parameters, merger-remnant characteristics, etc., with unprecedented precision~\cite{LIGOScientific:2020kqk, KAGRA:2021duu}. Among the different stages of binary black hole coalescence, ringdown represents the final phase in which the remnant black hole relaxes to a stationary Kerr state by emitting gravitational radiation~\cite{Regge:1957td, Vishveshwara:1970zz, Zerilli:1974ai, Teukolsky:1973ha, Berti:2005ys}. This phase is dominated by quasi-normal modes, where the emitted waveform is expressed as a linear superposition of oscillatory damped sinusoids. In general relativity, the no-hair theorem~\cite{Israel:1967za, Carter:1971zc, Robinson:1975bv} ensures that the quasi-normal mode frequency and damping time are functions of the mass and spin of the final Kerr black hole~\cite{Johannsen:2010ru, Meidam:2014jpa, Isi:2019aib, CalderonBustillo:2020rmh}. The quasi-normal mode encodes crucial information about the source properties~\cite{Berti:2005ys, Carullo:2019flw, Isi:2021iql, Cotesta:2022pci, Siegel:2023lxl}, spacetime geometry~\cite{Ghosh:2024het}, the underlying theory and potential deviation from general relativity~\cite{Isi:2020tac, Carullo:2021oxn, Mishra:2021waw, Pacilio:2023mvk, Silva:2022srr, Dey:2022pmv, Mishra:2023kng, Cano:2024ezp}. By analyzing the quasi-normal mode content in the ringdown data, black hole spectroscopy has emerged as a powerful tool to probe the intrinsic properties of black holes and the nature of compact objects. For an excellent review on both theoretical and observational aspects of black hole spectroscopy, including recent advancements in the subject, we refer the reader to Ref. \cite{Berti:2025hly}.

To extract such precise information from noisy data, Bayesian inference is widely used as the most optimal and accurate method. This approach involves the computation of a likelihood function that quantifies the probability of obtaining the data given a set of source parameters~\cite{Ashton:2018jfp, Thrane_2019, Roulet:2024cvl}. In the frequency domain, the likelihood is defined using the noise power spectral density (PSD), while in the time domain, it is obtained from the noise covariance matrix~\cite{Isi:2021iql, pyRing, Mishra:2023kng}. The posterior distribution is then obtained from the likelihood and a prior using various sampling algorithms such as Nested Sampling~\cite{Speagle_2020, john_veitch_2022_6460935} or Markov Chain Monte Carlo (MCMC)~\cite{2013ascl.soft03002F, pymc3}. In ringdown studies, where a specific portion of the data needs to be analyzed, the time domain approach is preferred as it allows direct access to the post-merger data without the transformation to the frequency domain. Although it avoids potential issues such as spectral leakage, edge effects, etc.,~\cite{Isi:2021iql}, it comes at the cost of computational efficiency. Moreover, with the inclusion of overtones and additional modified gravity parameters in the ringdown waveform, the prior space significantly expands, which further reduces the efficiency. 

Given such computational challenges in ringdown analysis, particularly with additional search parameters and an extended prior volume, it is important to explore other likelihood-free alternatives. Machine learning is one such promising framework that has been applied to a wide range of research problems in gravitational wave astronomy in recent years. This includes signal detection~\cite{Gabbard:2017lja, Murali:2022sba, Zelenka:2024vug, Wang:2023lif}, noise reduction~\cite{Ormiston:2020ele, Vajente:2019ycy, Yu:2021swq}, glitch identification~\cite{Wei:2019zlc, Torres-Forne:2020eax, Mogushi:2021cpw, Bacon:2022lsm}, waveform modeling~\cite{Khan:2020fso, GramaxoFreitas:2024bpk}, parameter inference~\cite{Gabbard:2019rde, Green:2020hst, Yamamoto:2020rse, Green:2020dnx, Dax:2021tsq, Bhagwat:2021kfa, Shen:2019vep, McLeod:2022ccr, Pacilio:2024qcq, Chatterjee:2024pbj, Bada-Nerin:2024wkn} etc. In this paper, we develop a Conditional Variational Auto-Encoder (CVAE) framework~\cite{NIPS2015_8d55a249, debbagh2025controllingstructuredoutputrepresentations} for accelerated ringdown parameter inference. A variational autoencoder (VAE)~\cite{kingma2022autoencodingvariationalbayes, Kingma_2019, doersch2021tutorialvariationalautoencoders} is a type of generative model that maps the input data to a latent space and then reconstructs it by sampling from a learned probability distribution. A CVAE extends this idea by feeding the network an additional conditioning input, which in our case is the corresponding true parameters. Note that a similar CVAE-based approach was previously explored in Ref.~\cite{Bhagwat:2021kfa}, focusing on simplified ringdown waveforms with fundamental modes only. In this work, we extend that framework by incorporating overtone excitations, mode phases, and beyond general relativity effects, enabling more complete and realistic black hole spectroscopy.

This approach enables us to analyze the ringdown data without relying on explicit likelihood evaluation. By training the neural network with simulated ringdown data from advanced LIGO detectors~\cite{LIGOScientific:2014pky, Capote:2024rmo}, we infer the posterior distribution of various key parameters. Our work is twofold. First, we apply the CVAE framework in the context of Kerr ringdown and obtain the posterior distribution of remnant mass and spin, along with amplitudes and phases. Furthermore, we extend this framework beyond general relativity by analyzing the ringdown of rotating black holes in braneworld theories. Specifically, we train a separate neural network to predict not only the Kerr parameters but also modified gravity parameters, such as the tidal charge. To further demonstrate the correctness of our results, we compare the posterior distribution obtained from the CVAE neural network with that of a Bayesian approach. 

The remainder of the paper is organized as follows. In \ref{sect:sec1}, we describe the ringdown waveform model and the procedure for generating synthetic training data. \ref{sect:sec2} introduces the CVAE neural network architecture, detailing its key components and the training algorithm. In \ref{sect:results}, we present the parameter inference results for both Kerr and braneworld scenarios. We compare the posterior distributions generated from the CVAE framework with those obtained from time-domain Bayesian analysis. We conclude in~\ref{sect:conclusion} by reviewing the key results and discussing possible future directions.

\begin{table*}[ht!]
\centering
\renewcommand{\arraystretch}{1.4}

\begin{minipage}[t]{0.3\textwidth}
\centering
\textbf{Encoder 1} \\[0.6em]
\begin{tabular}{|c|}
\hline
\textbf{Input:} $y \in \mathbb{R}^{410 \times 2}$ \\
\hline
\textbf{1D Conv. Layers} \\
\begin{tabular}{ccc}
Filters\,\,\, & Kernels \,\,\, & Strides \\
\hline
96 & 48 & 2 \\
96 & 32 & 2 \\
96 & 24 & 1 \\96 & 16 & 2 \\
96 & 8  & 1 \\
\end{tabular} \\
\hline
\textbf{Flatten}  \\
\hline
\textbf{FC Layers} \\
\begin{tabular}{c}
Neurons \\
\hline
2048 \\
1024 \\
512 \\
\end{tabular} \\
\hline
\textbf{Output}\\
$\mu_{l_d \times M_d}$, $\log \sigma_{l_d \times M_d}$, $w_{M_d \times 1}$ \\
\hline
\end{tabular}
\end{minipage}%
\hspace{0.03\textwidth}
\begin{minipage}[t]{0.3\textwidth}
\centering
\textbf{Encoder 2} \\[0.6em]
\begin{tabular}{|c|}
\hline
\textbf{Input:} $y \in \mathbb{R}^{410 \times 2}$ \\
\hline
\textbf{1D Conv. Layers} \\
\begin{tabular}{ccc}
Filters\,\,\, & Kernels\,\,\, & Strides \\
\hline
96 & 48 & 2 \\
96 & 32 & 2 \\
96 & 24 & 1 \\
96 & 16 & 2 \\
96 & 8  & 1 \\
\end{tabular} \\
\hline
\textbf{Flatten + Append ($x^{n}_{true}$)}  \\
\hline
\textbf{FC Layers} \\
\begin{tabular}{c}
Neurons \\
\hline
2048 \\
1024 \\
512 \\
\end{tabular} \\
\hline
\textbf{Output}\\
$\mu_{l_d \times 1}$, $\log \sigma_{l_d \times 1}$ \\
\hline
\end{tabular}
\end{minipage}%
\hspace{0.03\textwidth}
\begin{minipage}[t]{0.3\textwidth}
\centering
\textbf{Decoder} \\[0.9em]
\begin{tabular}{|c|}
\hline
\textbf{Input:} $y \in \mathbb{R}^{410 \times 2}$ \\
\hline
\textbf{1D Conv. Layers} \\
\begin{tabular}{ccc}
Filters\,\,\, & Kernels\,\,\, & Strides \\
\hline
96 & 48 & 2 \\
96 & 32 & 2 \\
96 & 24 & 1 \\
96 & 16 & 2 \\
96 & 8  & 1 \\
\end{tabular} \\
\hline
\textbf{Flatten + Append ($z_{l_d \times 1}$)}  \\
\hline
\textbf{FC Layers} \\
\begin{tabular}{c}
Neurons \\
\hline
2048 \\
1024 \\
512 \\
\end{tabular} \\
\hline
\textbf{Output}\\ $\mu_{n \times 1}$, $\log \sigma_{n \times 1}$ \\
\hline
\end{tabular}
\end{minipage}%
\caption{Each component (\texttt{Encoder 1}, \texttt{Encoder 2} and \texttt{Decoder}) is constructed using a stack of 1D CNN followed by a series of dense layers. Every convolutional layer is followed by $L_2$ regularization (rate $0.001$), batch normalization and a LeakyReLU activation. The output of the final convolutional block is flattened before passing to the dense layers. In \texttt{Encoder 1}, the flattened feature map is directly passed through three dense layers, each followed by batch-normalization, LeakyReLU activation and dropout. The flattened output of \texttt{Encoder 2} is concatenated with the true parameters ($x^n_{true}$) before passing through the dense layers. In the \texttt{Decoder}, the flattened object is concatenated with the \texttt{Encoder 2} latent sample ($z_{l_d \times 1}$)  and the combined representation is passed through the dense layers. To prevent overfitting, we apply dropout with a rate of $0.2$ after each dense layer. The final output of the decoder represents the mean and log variance corresponding to all search parameters ($n = 6$ and $7$ for Kerr and braneworld). The means are passed through a sigmoid activation, whereas for the log variance, we use negative ReLU.}
\label{tab:nn_arch}
\end{table*}

\section{Ringdown Waveform \& Data Preparation}\label{sect:sec1}
After the merger of two black holes, the remnant object undergoes a relaxation phase known as the ringdown. This phase is dominated by the system's characteristic quasi-normal modes that govern how the perturbations are radiated as gravitational waves. The quasi-normal modes are solutions to the linearized Einstein's equation with ingoing and outgoing boundary conditions at the event horizon and spatial infinity. These modes are characterized by complex frequencies with the real part corresponding to the oscillation frequency and the imaginary part dictating the damping rate. Gravitational radiation in the ringdown phase can be modelled as a linear superposition of these quasi-normal modes~\cite{Berti:2005ys},
\begin{align}\label{waveform}
h_{+}(t)&+i \, h_{\times}(t)=\sum_{lmn}\Big[\mathcal{A}_{lmn} e^{(i\,\omega_{lmn} - 1/\tau_{lmn}) (t - t_0)} 
\nonumber
\\
&\qquad\qquad \times S_{lmn}(\iota, \phi) e^{i\, \phi_{lmn}} \Big]~.
\end{align}
Here $(\mathcal{A}_{lmn}, \phi_{lmn})$ denotes the amplitude and phase corresponding to the quasi-normal mode $(l,m,n)$, with $n$ being the overtone number and $(l, m)$ denoting the angular indices. The angular contribution to the waveform comes from the spin-weighted spheroidal harmonics $S_{lmn}(\iota, \phi)$, where $(\iota, \phi)$ are inclination and azimuthal angle. The most crucial components of the waveform are the frequency ($ \omega_{lmn} = 2\pi\,f_{lmn}$) and damping time ($\tau_{lmn}$). Note that we neglect the contribution of the counter-rotating (retrograde) modes, as they are typically weakly excited in aligned-spin binaries~\cite{Berti:2005ys, Lim:2019xrb, Li:2021wgz}. The strain time series measured at the detector is expressed as the linear combination of the two fundamental polarization states, modulated by the detector's response,
\begin{equation}\label{projection}
h(t) = \sum_{ i = +, \times} F_{i}(\alpha, \delta, \psi), h_{i}(t)~.
\end{equation}
where $F_{i}$ represents the detector antenna pattern functions, which depend on the sky localization parameters right ascension $(\alpha)$, declination ($\delta$) and polarization angle ($\psi$). In general relativity, the black hole no-hair theorem ensures that the quasi-normal modes are uniquely determined from the final mass and spin of the remnant Kerr black hole. However, this is no longer true in the presence of beyond general relativity effects, as additional parameters may influence the quasi-normal mode spectrum. In particular, for the braneworld theory~\cite{Maartens:2003tw, Perez-Lorenzana:2005fzz}, in the presence of a tidal charge parameter ($q$), we have : $f_{lmn} = f_{lmn}(M, \chi, q)$ and $\tau_{lmn} = \tau_{lmn}(M, \chi, q)$. 


\begin{figure}[!tb]
\centering
\includegraphics[width=0.48\textwidth]{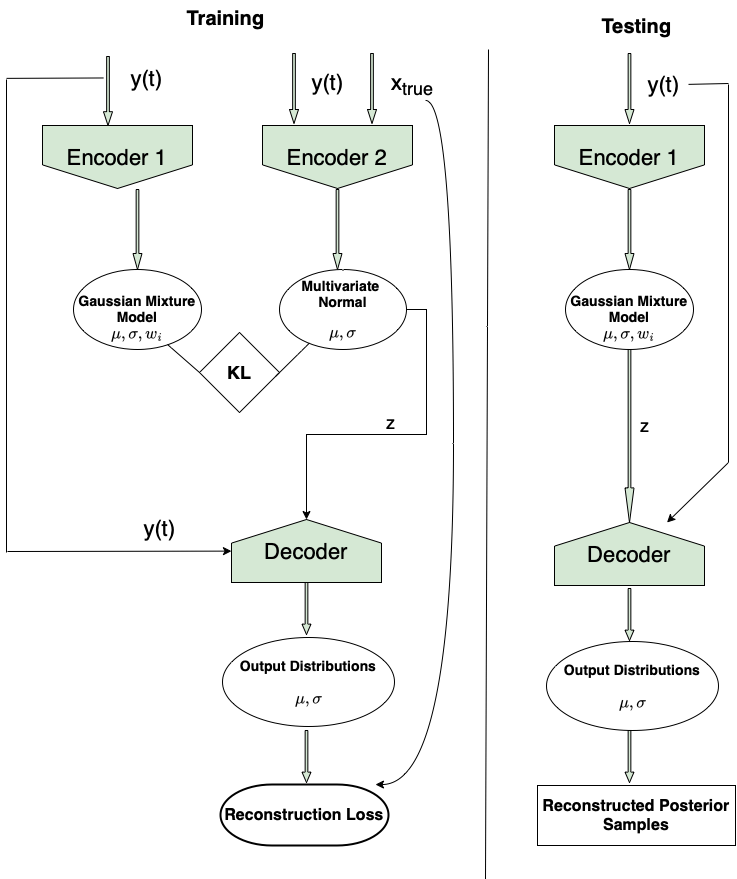}
\caption{A visual overview of the training and testing algorithm. The left panel illustrate a single training step, while in the right panel we show the inference procedure on a test dataset. See main text for a detailed explanation.}
\label{training_flowchart}
\end{figure}

To build the training dataset, we generate simulated ringdown waveforms using \ref{waveform} and \ref{projection}, projecting signals onto the Hanford and Livingston detectors. For Kerr quasi-normal modes, we use the fitting formulas from Ref. \cite{Berti:2005ys}, while for the braneworld case, we follow the approach outlined in Ref. \cite{Mishra:2021waw, Mishra:2023kng}. The injection parameters are sampled uniformly from astrophysically relevant intervals. For the Kerr case, the prior ranges are taken to be: $M_f \in [20, 150] M_\odot$, $\chi \in [0.01, 0.99]$, $A_{22n} \in [10^{-22}, 10^{-20}]$, $\phi_{22n} \in [0, 2\pi]$. In the braneworld scenario, the additional tidal charge parameter $q \in [0.0, -1.0]$, and the spin assumes a dynamical prior given by $\chi \in [0.0, \sqrt{1 - q}]$~\cite{Mishra:2023kng}. For simplicity, we keep the sky locations fixed: $(\alpha, \delta) = (1.95, -1.22)$. We also work with fixed values of polarization, inclination, and azimuthal angle, i.e. $(\psi, \iota, \phi) = (0.82, \pi, 0.0)$.  Each ringdown signal is 0.1 seconds long, with the ringdown start time fixed at the merger ($t_0 = 0.0$) and a sampling rate of $4096$ Hz. We consider the fundamental mode and one additional overtone ($n = 1$)in the waveform. 
The waveforms are then whitened with the detector's noise covariance matrix, which is used as the base input during training. We restrict the training dataset to signals having a minimum SNR of $15$.

\begin{figure*}[!tb]
\centering
\includegraphics[width=0.49\textwidth]{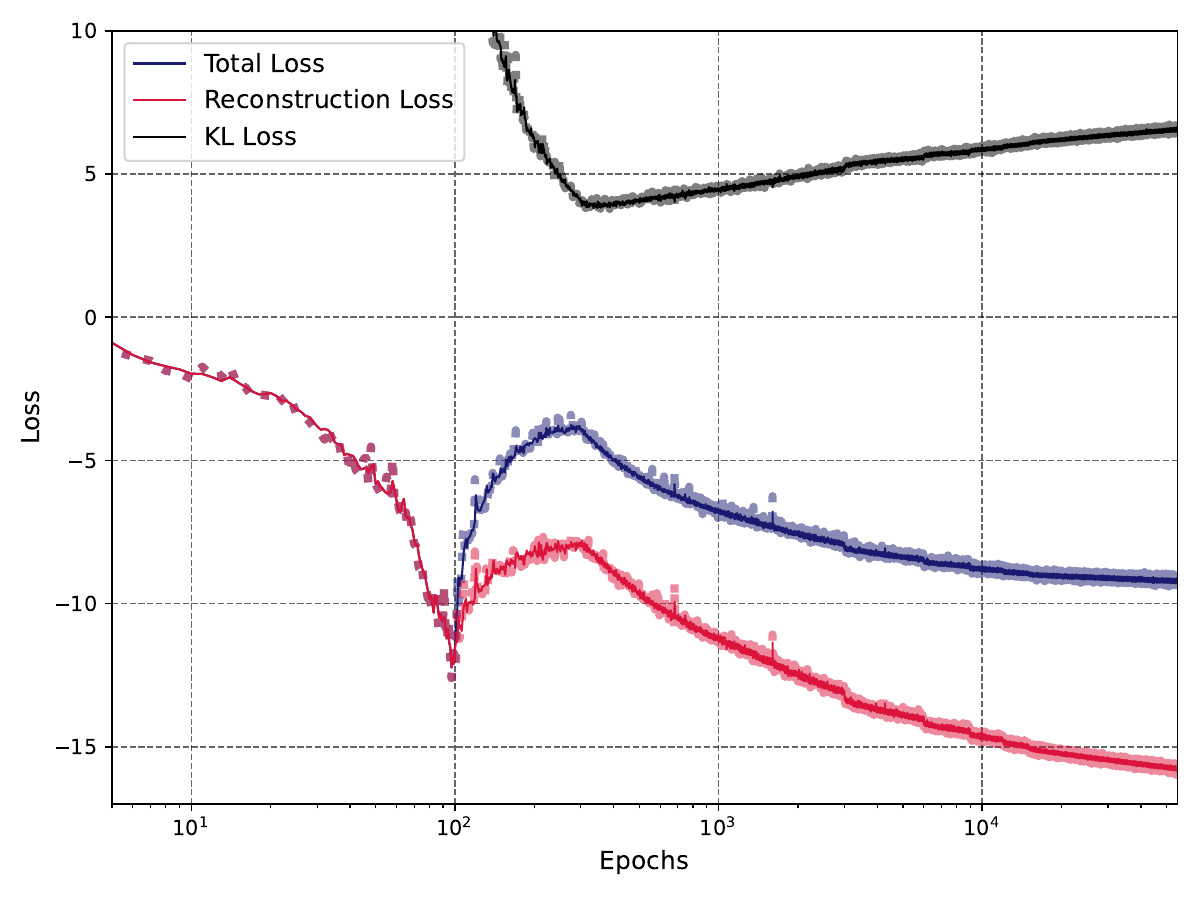}
\includegraphics[width=0.49\textwidth]{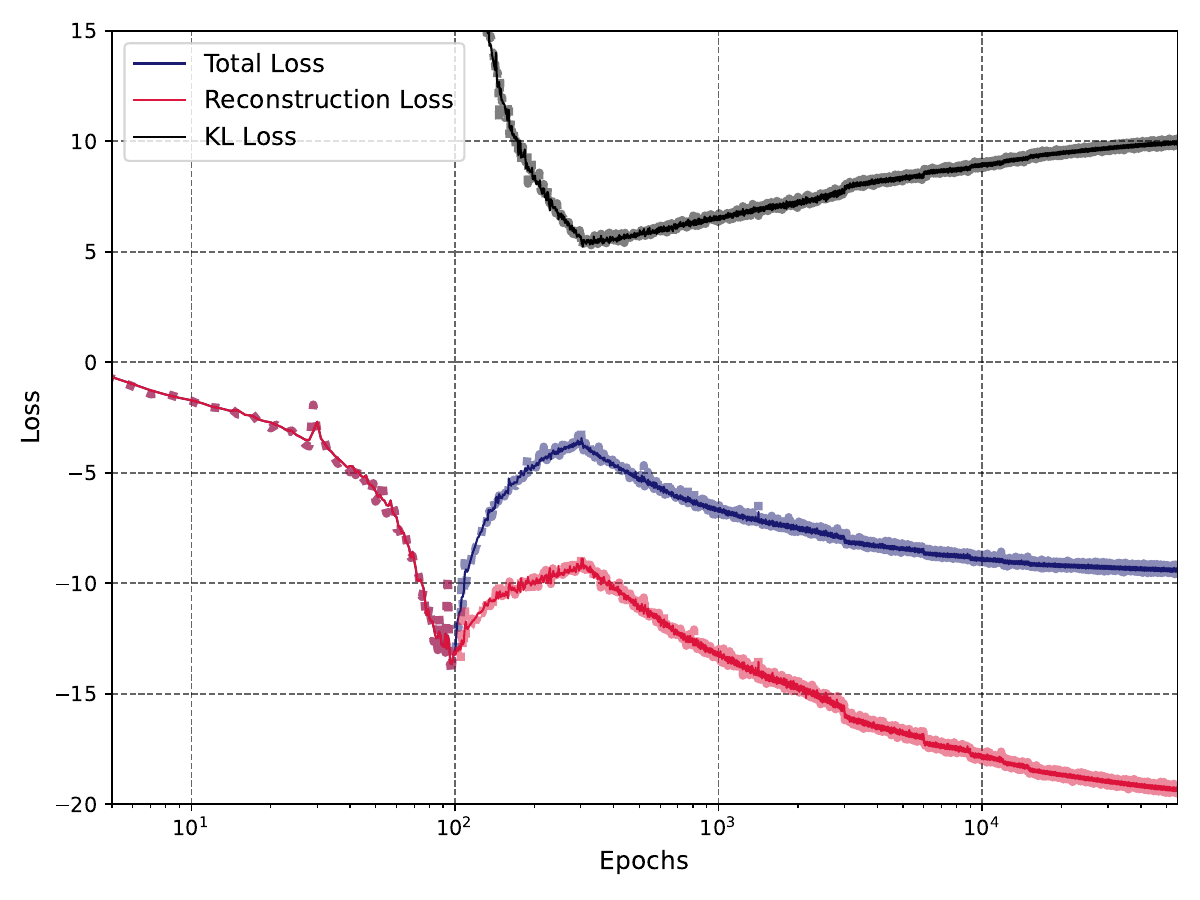}
\caption{
This figure shows the evolution of the loss components as a function of training epochs for both the Kerr (left panel) and the braneworld (right panel) models. We display the KL divergence (black), reconstruction loss (red) and total loss (blue) for training and validation.}
\label{loss_plot}
\end{figure*}

\section{Neural Network Architecture and Training Algorithm}\label{sect:sec2}
Having briefly described the ringdown waveform and training sample generation in the previous section, we now turn our attention to the CVAE neural network architecture and associated training algorithm. We closely follow the structure and algorithm of Ref. \cite{Gabbard:2019rde}. The CVAE network is comprised of two encoders and a single decoder. The first encoder (\texttt{Encoder 1}) processes the input data $y(t)$ into a lower dimensional latent space representation modelled by a Gaussian Mixture Model (GMM). The output of \texttt{Encoder 1} comprises the complete set of parameters representing the GMM: means, variances, and weights corresponding to various mixture components. We work with a latent space of dimension ($l_d$) $16$ and the number of mixture components ($M_d$) of the GMM is taken to be $24$. The second encoder (\texttt{Encoder 2}) takes both the time series data and associated true parameters ($x_{true}$) as input to learn a latent representation modelled by a Multivariate Normal (MVN) distribution. The output of \texttt{Encoder 2} consists of mean and variance vectors in the latent space corresponding to the MVN distribution. Note that, we use \texttt{Encoder 2} as an auxiliary component during the training that will be discarded during inference. The first component of our loss function comes from the Kullback–Leibler (KL) divergence between the latent distribution produced by \texttt{Encoder 1} and \texttt{Encoder 2}. 

The \texttt{Decoder} takes the time series data along with a latent sample $z$ drawn from the \texttt{Encoder 2} distribution as input and returns the mean and variance corresponding to the physical parameters. The \texttt{Decoder}'s output distribution for the final mass, final spin and amplitudes are modelled by truncated Gaussians, which ensures the parameters remain within physically meaningful bounds. In contrast, phases being angular and periodic are modelled using a Von-Mises distribution. The final component of our loss function is the reconstruction loss between the output of the \texttt{Decoder} and true parameters, which is computed as the negative log probability of the true parameters over the output distribution. The final loss is given by the sum of these two contributions~\cite{Gabbard:2019rde}. 
\begin{equation}
    \mathcal{L}_{total} = \mathcal{L}_{Reconstruction} + \beta\, \mathcal{L}_{KL}
\end{equation}
where $\beta$ is a dynamically evolving factor that follows a log-scheduler, which progressively increases from $0$ to $1$ between epochs $100$ and $300$ during the training to gradually introduce the KL divergence component. The network is trained to minimize $\mathcal{L}_{total}$. In \ref{training_flowchart}, we illustrate the training and inference algorithms.

Both encoders and decoder consist of a sequence of one-dimensional convolutional neural network (CNN) layers followed by fully connected (FC) layers. The architectural details of the neural network, which include the number of 1D-CNN layers, number of filters, kernels, strides, and neurons etc, are summarized in \ref{tab:nn_arch}. During the training, we tune various key hyper-parameters such as the number of filters, kernel size, strides corresponding to each 1D CNN layer, number of neurons in the FC layers, dropout rate, learning rate, etc. The training dataset consists of approximately $3$ million samples. In our training scheme, we identify a training epoch as the instance where the network has been trained on $16 \times 1024$ samples. We load a new training dataset of $16 \times 1024$ whitened waveform samples every four epochs. At each load, we add zero-mean, unit-variance Gaussian noise, ensuring the same training samples are exposed to different noise realizations.  We use the Adam optimizer with an initial learning rate of $10^{-4}$, which is reduced by $25\%$ every $3000$ epochs until $15000$ epochs, after which it remains constant.  The training is performed with a batch size of $1024$ and continued for $55000$ epochs, taking approximately $40$–$45$ hours on a single GPU. Despite the increased complexity of the parameter space due to the inclusion of the tidal charge parameter, the braneworld network is trained using the same hyperparameter configuration as the Kerr network. We present the loss function for both Kerr and Braneworld models in \ref{loss_plot}. The close agreement between the training and validation losses suggests that the network is not overfitting to the training dataset. 
\begin{figure*}[!tb]
\centering
\includegraphics[width=0.49\textwidth]{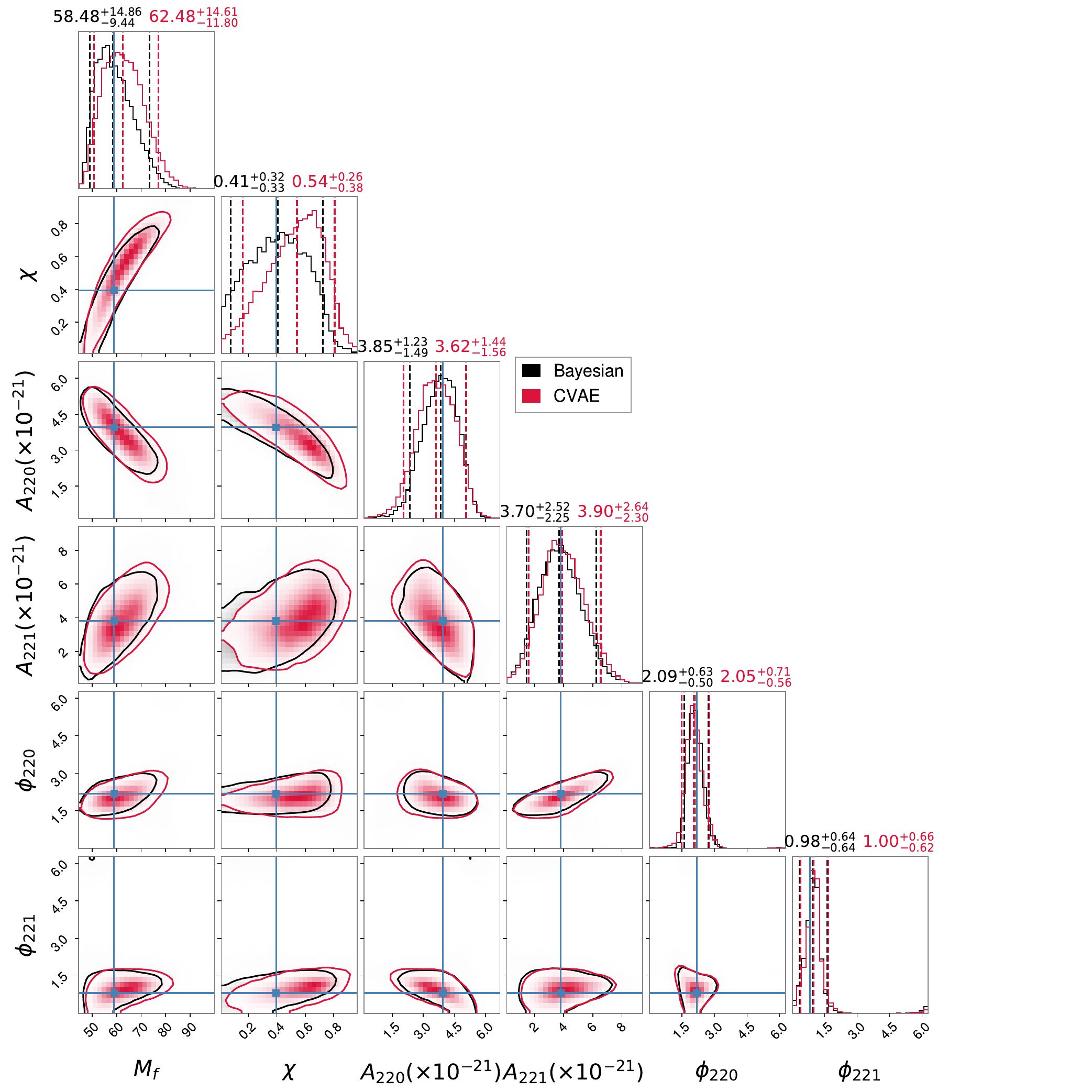}
\includegraphics[width=0.49\textwidth]{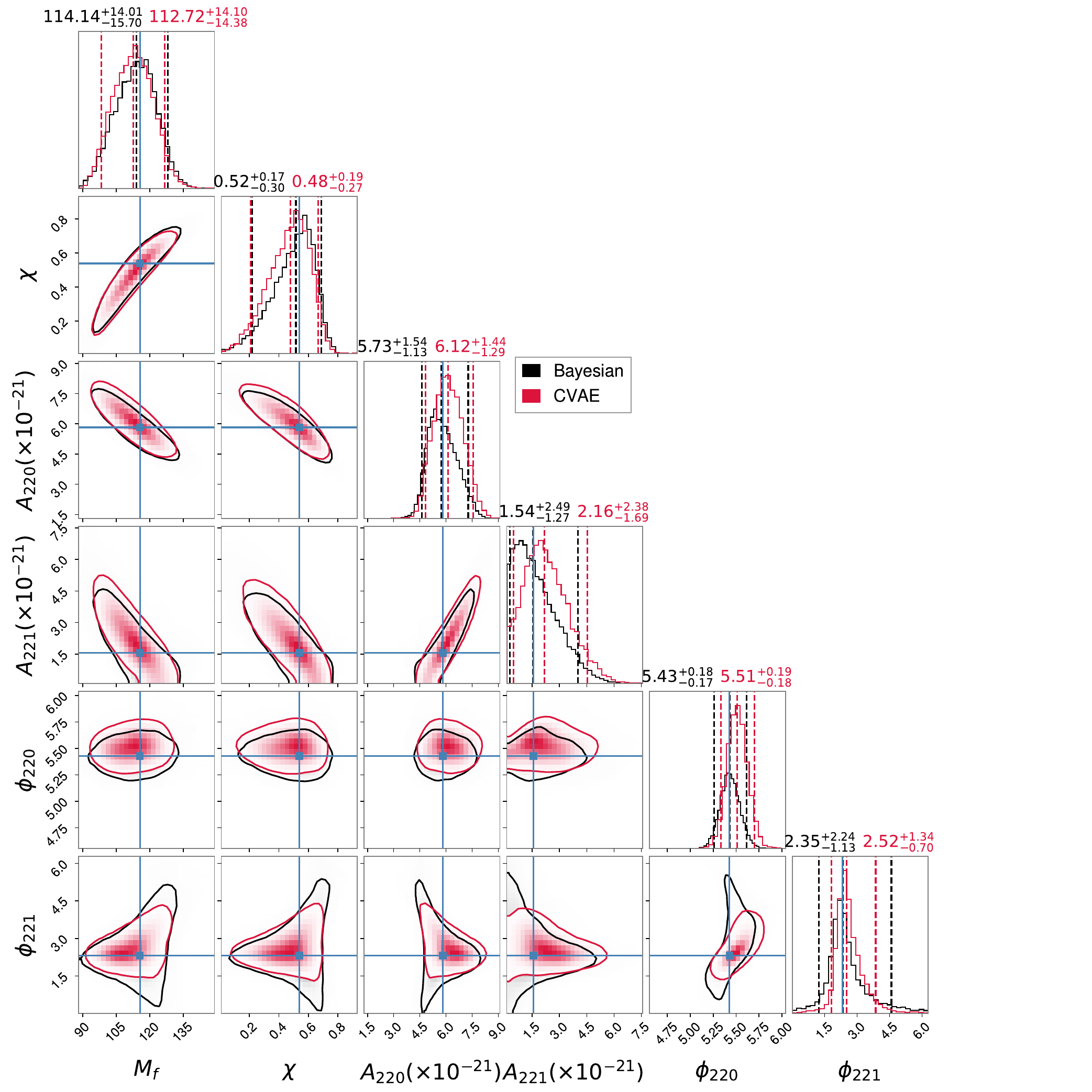}
\includegraphics[width=0.49\textwidth]{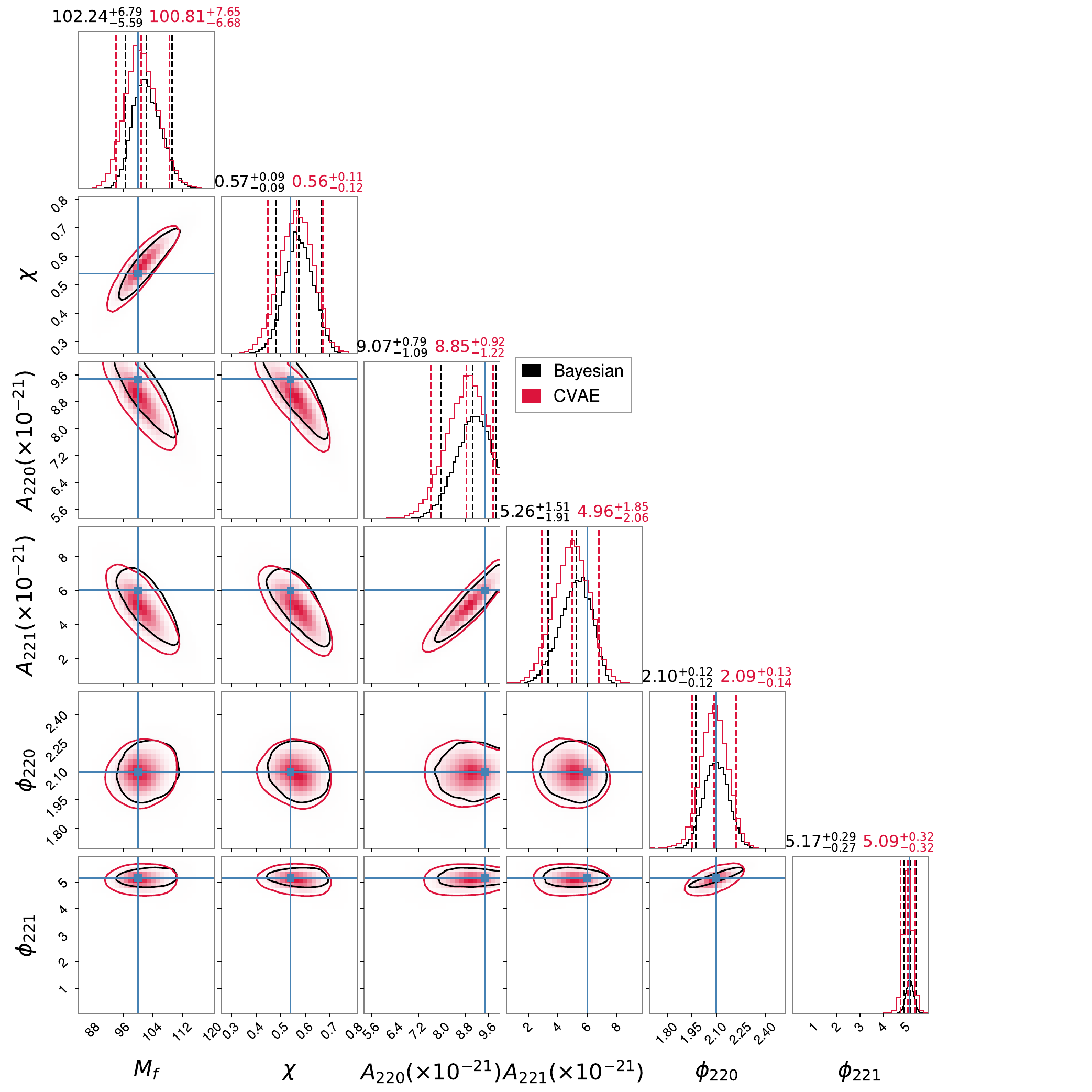}
\includegraphics[width=0.49\textwidth]{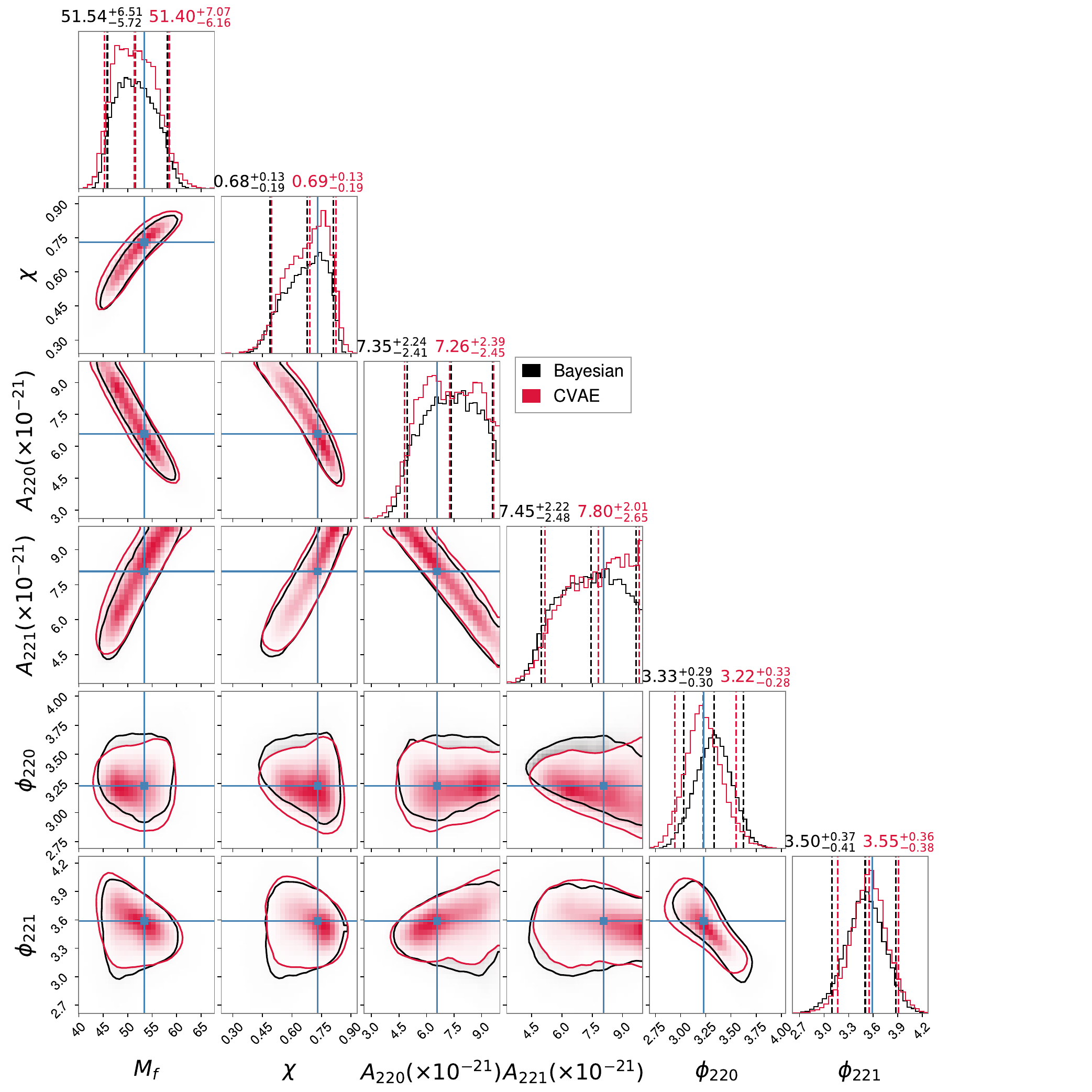}
\caption{
In this figure, we present the CVAE constructed posterior (crimson) distribution and compare it with that of the Bayesian inference results (black). The top left panel corresponds to a signal injection with an SNR of $\sim 30$, whereas the top right panel corresponds to an SNR of $\sim 45$. The bottom left and right panel shows the same comparison with a higher SNR injection of $\sim 60$ and $85$, respectively. This side-by-side comparison demonstrates the performance of the neural network across different SNRs. 
}
\label{kerr_inference}
\end{figure*}

\section{Results}\label{sect:results}
In this section, we examine the trained CVAE model on a variety of test datasets corresponding to different black hole configurations and gravity theories. The objective is to assess the model's accuracy in predicting physical parameters of the remnant black hole. Once training is complete, \texttt{Encoder 2} is discarded, and inference is performed using only \texttt{Encoder 1} and the \texttt{Decoder}. The test sample is first encoded into a latent representation via \texttt{Encoder 1}. Samples drawn from this latent space are then combined with the ringdown data and passed through the \texttt{Decoder} to produce output distributions corresponding to various search parameters. 
\subsection{Kerr}
\begin{figure*}[!tb]
\centering
\includegraphics[width=0.49\textwidth]{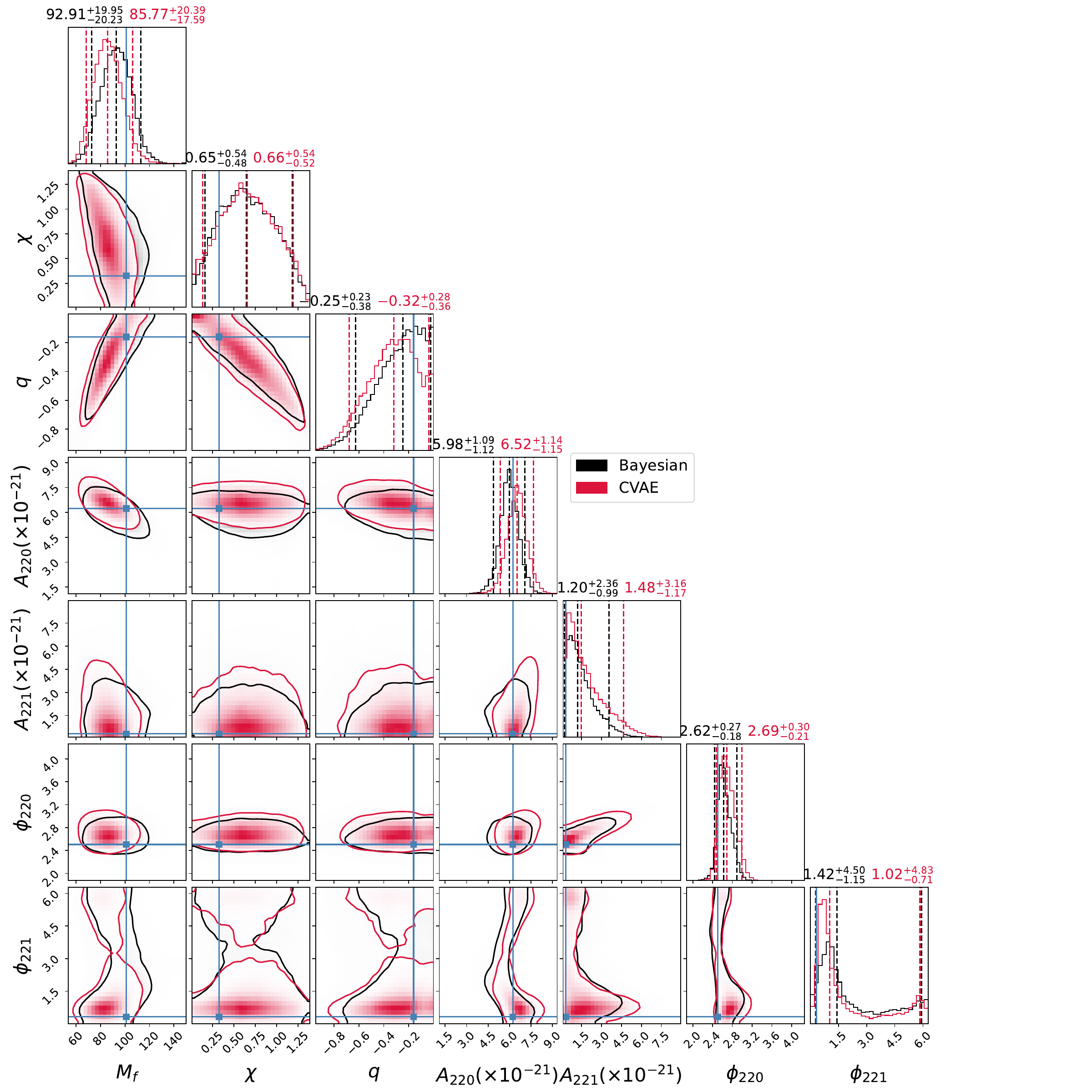}
\includegraphics[width=0.49\textwidth]{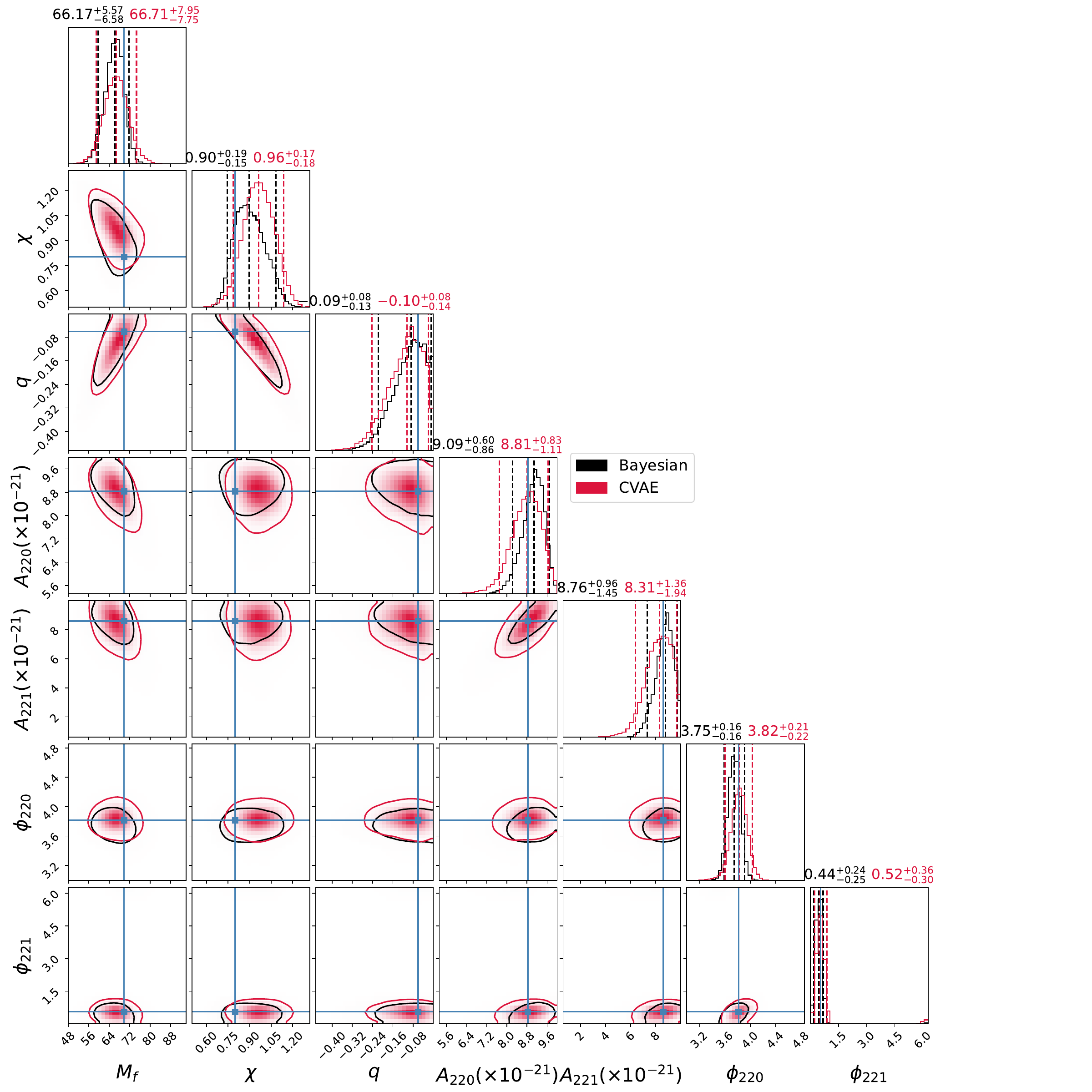}
\caption{
In this figure, we present the CVAE constructed posterior (crimson) distribution and compare it with that of the Bayesian inference results (black) for the braneworld model. The left panel corresponds to a signal injection with an SNR of $\sim 30$, whereas the right panel corresponds to an SNR of $\sim 45$.
}
\label{tidal_inference}
\end{figure*}
\begin{figure*}[!tb]
\centering
\includegraphics[width=0.49\textwidth]{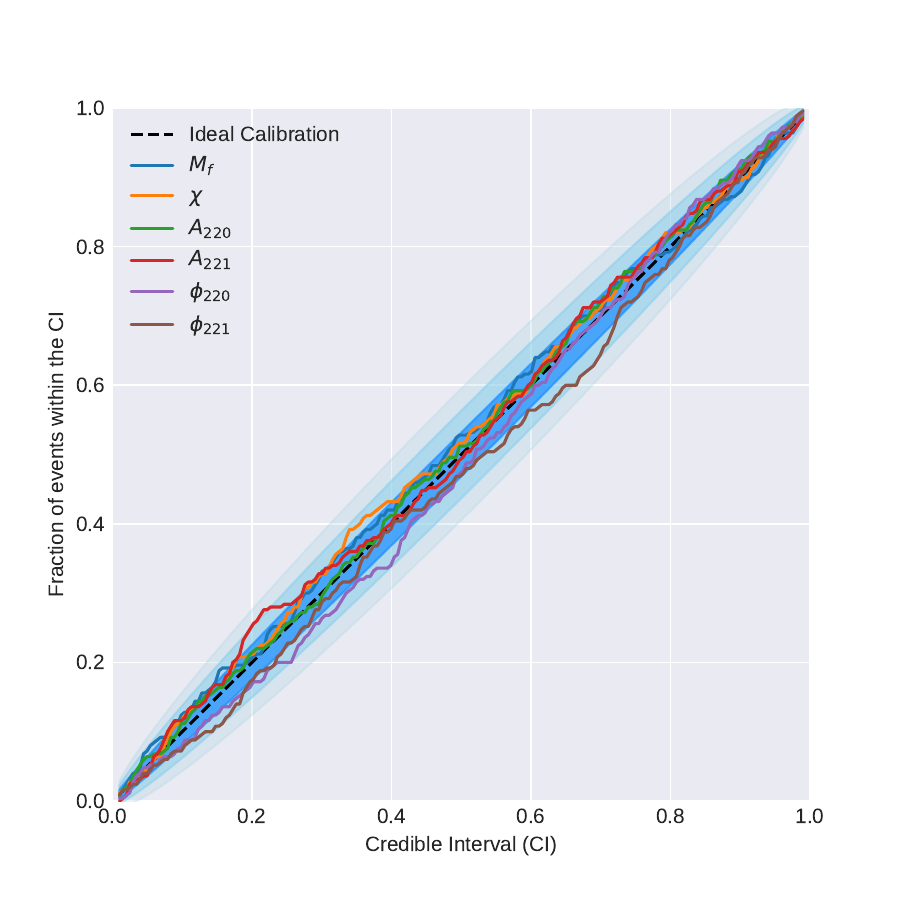}
\includegraphics[width=0.49\textwidth]{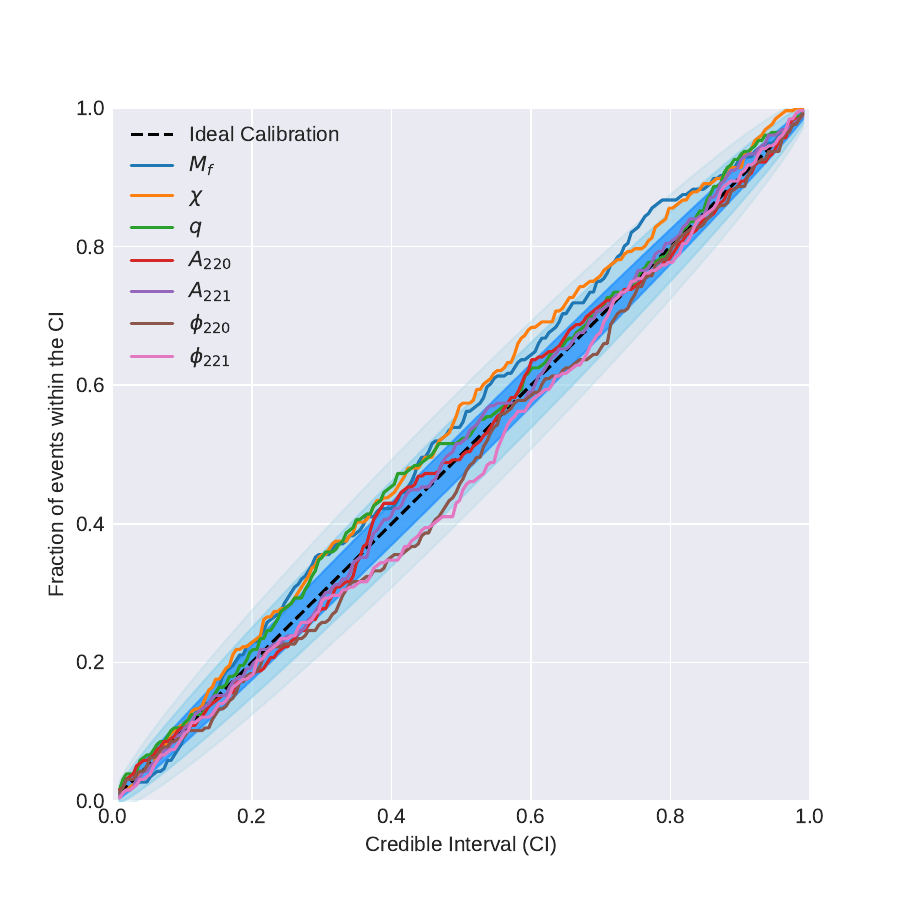}
\caption{
In this figure, we show the PP plot for Kerr inference (left panel) and braneworld inference (right panel) using the neural network on the test dataset, obtained using 256 test samples. The $x$-axis denotes the nominal confidence level (i.e., the $p\%$ credible interval), and the $y$-axis shows the fraction of true values contained within that interval. A well-calibrated posterior should closely follow the diagonal line (ideal calibration). The Kolmogorov–Smirnov (KS) test p-values (listed top-to-bottom as in the PP plot) are
$[0.73,0.59,0.67,0.33,0.26,0.17]$ for Kerr, and  $[0.29, 0.36, 0.47, 0.65, 0.58, 0.21, 0.14]$ for the braneworld case, indicating statistically consistent calibration. The shaded regions indicate the $1\sigma,\, 2\sigma,\, 3\sigma$ confidence level.
}
\label{pp_plot}
\end{figure*}

We begin by evaluating the performance of our CVAE model on a test dataset generated by assuming the remnant to be a Kerr black hole. The test dataset includes time series sampled from a wide range of final mass, spin, amplitudes and phases drawn from astrophysically motivated prior distributions. The neural network is tasked to approximate the posterior distribution over these search parameters from the input time series data. To further validate our results, we compare the neural network predicted posterior with those obtained through standard Bayesian inference. For the time domain Bayesian parameter estimation we follow the sampling framework developed in \cite{Mishra:2023kng} using \texttt{dynesty}. In particular, we employ the \texttt{NestedSampler} with \texttt{‘unif’} sampling method and $3000$ live points to ensure well-converged posteriors. In \ref{kerr_inference}, we present a direct comparison between the posterior distribution obtained from the CVAE neural network and Bayesian inference. This figure illustrates the corner plot over the posterior distributions of all search parameters at various SNRs. As evident, the CVAE-generated posterior shows strong agreement with that of the Bayesian inference results, demonstrating the efficacy of our neural network. To further assess the calibration of our CVAE-generated posterior samples, we present a probability–probability (PP) plot in Fig. \ref{pp_plot} (left panel). For a well-calibrated model, the PP curves should closely follow a diagonal line.

\subsection{Braneworld}
We next evaluate the performance of our CVAE neural network on a dataset generated assuming the Braneworld hypothesis. The ringdown waveforms now includes an additional deviation parameter, the tidal charge. To evaluate the trained network performance, we again compare the CVAE output posterior distribution with the standard time-domain Bayesian inference obtained using \texttt{dynesty}. As shown in \ref{tidal_inference}, the CVAE posteriors shows excellent agreement with the Bayesian benchmarks. This demonstrates that the network is able to reliably capture the effects of modified gravity deviations in the signal. The corresponding PP plot is presented in Fig. \ref{pp_plot} (right panel).

Unlike the time-domain Bayesian inference, which can be computationally expensive, the CVAE framework offers a significant speed advantage once the model is trained. For instance, generating $20000$ posterior samples for a given ringdown signal takes only about $4-5$ seconds. This represents a gain of several orders of magnitude in efficiency, making such machine learning based analysis well-suited for real-time and large-scale data analysis. 

However, despite the overall strong performance of the neural network, there are still some limitations worth noting. While the neural network demonstrates good agreement with Bayesian results across a broad range of parameter space, its performance degrades when multiple injected parameters simultaneously lie close to the prior edge. This is likely due to reduced training sample density in those regions. Additionally, we also find that the CVAE neural network performance reduces for low SNR injections ($\sim <20$), whereas for medium and high SNR, the predicted posteriors are more accurate. These limitations are likely addressable in future with rigorous training algorithms, emphasizing sampling near the prior boundaries and improved hyperparameter optimization.

\section{Conclusion}\label{sect:conclusion}
Motivated by increasing complexity and computational inefficiency in traditional time-domain Bayesian analysis with higher-dimensional prior space (either due to the presence of higher-order overtones or additional hairs), we developed a machine learning approach capable of rapidly generating posterior distributions over the search parameters. In particular, we have presented a CVAE framework for efficient and likelihood-free parameter estimation of ringdown from gravitational wave observation. Once the neural network is trained, inference can be performed within seconds, in contrast to the Bayesian approach, which typically requires several hours to complete. Firstly, we implemented the CVAE framework in the context of perturbed Kerr black holes and extracted key physical parameters such as remnant mass, spin, amplitudes, and phases. Further, we extended this model to analyze deviation from general relativity and particularly studied the ringdown of perturbed black holes in braneworld theories. Black holes in such theories are charactrized by a non-zero tidal charge parameter in addition to the remnant mass and spin. We trained a separate neural network that can capture such deviation from general relativity and infer the additional parameters. Our result demonstrates that neural networks such as CVAE offer a computationally efficient alternative to likelihood-based Bayesian inference for black hole spectroscopy. In both of these cases, we have compared the posterior distribution obtained from the CVAE neural network with the results of Bayesian inference and found good agreement. 

While the CVAE neural network offers a significant speed advantage in producing reliable posterior distributions compared to Bayesian approaches, certain limitations persist, particularly in low-SNR regimes and near the edges of the prior space. We aim to address these challenges in future work. Looking ahead, our framework can be extended to include more complex waveform models, real detector noise, non-Gaussianity, or additional degrees of freedom from modified theories of gravity. We aim to extend the inference to include sky localization parameters and ringdown start time in subsequent works.\\

\section{Acknowledgment}
I am grateful to Arunava Mukherjee, Prayush Kumar, Shilpa Kastha, and Emanuel Hoque for useful discussions. I especially thank Gregorio Carullo and Lalit Pathak for their detailed reviews of the manuscript and valuable suggestions. I extend my gratitude to the members of the “Parameter Estimation” group within the LIGO–Virgo–KAGRA collaboration where the work was presented. The computations were performed using an NVIDIA Tesla V100-PCIE-16GB GPU on the CCS2 cluster at SINP. The content of this manuscript has been derived using the publicly available \texttt{python} software packages: \texttt{dynesty, lalsuite, matplotlib, numpy, scipy, optuna, tensorflow, corner}.

\bibliography{Journal_V1}
\bibliographystyle{./utphys1}
\end{document}